\documentclass{ws-acs}
\usepackage{url}
\usepackage{graphicx}
\usepackage{amssymb,amsmath}

\begin{document}

\makeatletter
\def\@biblabel#1{[#1]}
\makeatother


%
\catchline{}{}{}{}{}
%

\title{Resource Redistribution Method \\ for Short-Term Recovery of Society after Large Scale Disasters}

\author{\footnotesize VASILY LUBASHEVSKIY,\footnote{e-mail: kloom@mail.ru} \quad
TARO KANNO,\footnote{e-mail: kanno@sys.t.u-tokyo.ac.jp}\quad
KAZUO FURUTA\footnote{e-mail: furuta@rerc.t.u-tokyo.ac.jp}}
\address{Department of Systems Innovations, School of Engineering, University of Tokyo,\\ 7-3-1 Hongo Bunkyo-ku,
Tokyo, 113-8656, Japan}

\maketitle

\begin{abstract}
Recovery of society after a large scale disaster generally consists of two phases, short- and long-term recoveries. The main goal of the short-term recovery is to bounce the damaged system back to the operating standards enabling residents in damaged cities to survive, and fast supply with vital resources to them is one of its important elements. We propose a general principle by which the required redistribution of vital resources between the affected  and neighboring cities can be efficiently implemented. The short-term recovery is a rescuer operation where uncertainty in evaluating the state of damaged region is highly probable. To allow for such an operation the developed principle involves two basic components. The first one of ethic nature is  the triage concept determining the current city priority in the resource delivery. The second one is the minimization of the delivery time subjected to this priority. Finally a certain plan of the resource redistribution is generated according to this principle. Several specific examples are studied numerically. It elucidates, in particular, the effects of system characteristics such as the city limit capacity in resource delivery, the type of initial resource allocation among the cities, the number of cities able to participate in the resource redistribution, and the damage level in the affected cities. As far as the uncertainty in evaluating the state of  damaged region is concerned,  some specific cases were studied. It assumes the initial communication systemhas crashed and formation of a new one and the resource redistribution proceed synchronously. The obtained results enable us to consider the resource redistribution plan governed by the proposed method  semi-optimal and rather efficient especially under uncertainty. 
\end{abstract}


\section{Introduction}
\label{23}

In recent years the problems of disaster mitigation and resilience have attracted much attention. As far as mitigation of large scale disasters is concerned, two phases, short- and long-term recoveries, can be distinguished. Use of these terms has a long history \cite{national1979comprehensive}, nonetheless, the appropriate classification of recovery phases is required especially for efficient emergency management of large scale disasters \cite{DHS2008,MalcolmBaird2010a,DHS2013}.   

Following the cited materials we consider the short-term recovery to be \textit{mainly} aimed at restoring the vital life-support system to the minimal operating standards required for surviving. Generally this system comprises many individual components and the corresponding services, in particular, sheltering, feeding operations, emergency first aid, bulk distribution of emergency items, and collecting and providing information on victims to their family members. One of the basic requirements imposed on the short-term recovery is  beginning  its implementation within the minimal time. For example, the aforementioned services have to start their operation within 8 hours according to the Disaster Recovery Plan of State of Illinois \cite{DRP_Illinois_July2012_STRecovery}.

To mitigate aftermath of a large scale disaster cooperation of many cities is required, because the amount of resources initially accumulated in an affected area can be insufficient to recover all the individual components of the vital life-support system. Thereby the implementation of the short-term recovery is directly related to an efficient resource redistribution. Although various early warning systems are in use, the precise prediction of the critical infrastructure damage is still a hard  problem especially in case of large scale disasters. Therefore the recovery implementation cannot be preplanned reliably and it is possible only to formulate rather general requirements for this process. First, supply to an affected area must start practically immediately in order to recover the life-support system. Second, the resource supply should be decentralized, otherwise, its centralized management can be a `bottleneck' that delays the responsive and adaptive delivery of resources or aid \cite{managprob:panda2012}.

According to literature the problem of resource redistribution for recovery from large-scale disasters is related to emergency or humanitarian logistics (for a recent review see, e.g., \cite{caunhye2012optimization}). A rather general and detailed formulation of the problem was given by Haghani and Oh \cite{haghani1996formulation}. It is based on a multi-commodity, multi-modal network flow model for disaster relief operations and generates routing and scheduling plans for multiple transportation modes carrying various relief commodities from multiple supply points to demand points in the disaster area. The pivot point of such approaches is minimizing the sum of  vehicular flow costs, commodity flow costs, supply/demand storage costs and inter-modal transfer costs over all time periods as the main objective. A similar concept was explored in \cite{lin2011logistics} with the same objective (cost optimization). A relative problem was studied in \cite{afshar2012modeling} where, however, the cost minimization was replaced by the condition of minimizing the total amount of weighted unsatisfied demand in the affected region.

The integral characteristics like the total cost or the total unsatisfied demand are natural measures of the efficiency of  long-term recovery. In case of short-term recovery the key objective is minimizing its implementation time. The total time required to finalize the short-term recovery process is not a linear functional of supply flows and, thus, the problem of its minimization is not described by the linear programming formalism.   

The purpose of the present paper is developing a method by which such resource redistribution can be implemented.  Two particular cases of the initial resource distribution, centralized and uniform, will be studied in detail. The supply dynamics depending on the number of elements in the suppling network and the city limit capacity will be investigated. Besides, it will be demonstrated that the method to be proposed enables efficient recovery even the information about the damaged region is gradually accumulated during the recovery process.

\section{Model}

\begin{figure}[t]
\begin{center}
\includegraphics[width=1.0\columnwidth]{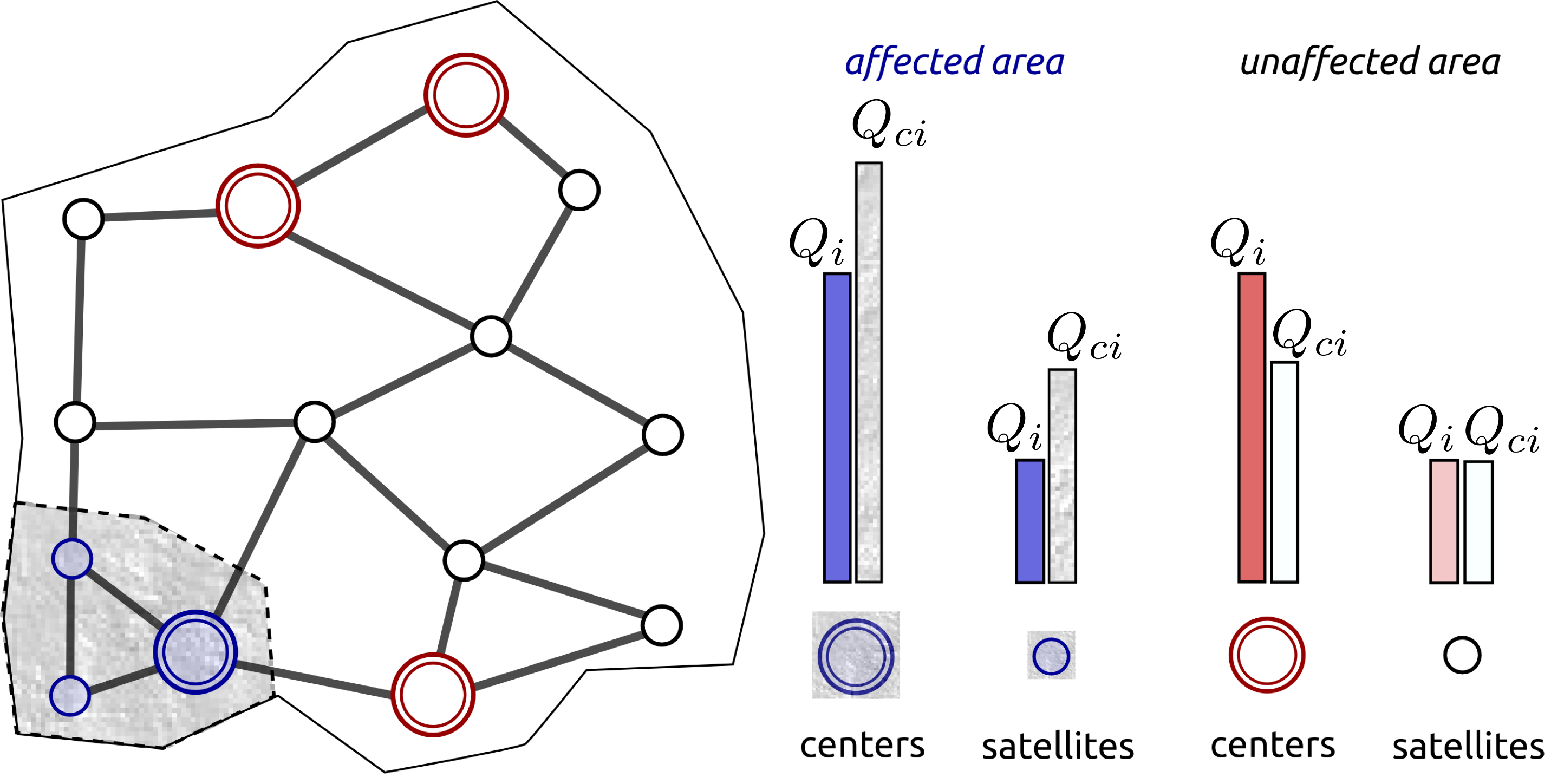}
\end{center}
\caption{A schematic illustration of the ``centralized'' system. The affected area is shadowed and for the cities there the minimal critical level $Q_{ci}$ becomes higher than the current amount of resources $Q_i$.}
\label{F1}
\end{figure}

\subsection{Model background}\label{modback}

The Great East Japan Earthquake occurred along the eastern coast of Japan on 11th of  March, 2011 exemplifies large scale destructive disasters that necessitate cooperation of many cities and even regions in mitigating the aftermath. The hypo-central region of this earthquake comprised several offshore prefectures (Iwate, Miyagi, Fukushima, and Ibaraki Prefectures) and have  ruptured the zone with a length of 500~km and a width of 200~km \cite{wiki:GEJ2011}. The terrible aftermath of the disaster initiated evacuation from some areas of these prefectures, thousands houses were destroyed, many victims required medical assistance. Obviously none of the affected cities was able to recover only by its local resources, practically all the non-affected cities in these prefectures were involved into this process. 
New shelters were urgently built in many non-affected regions, some highways were closed for private vehicles, flows of required pure water, food, medical drugs, fuel, etc. was redirected to the damaged cities. 
The ability to modify urgently the supplying system is one of the crucial points for a high resilience of the system as a whole.
These Japanese prefectures can be one of the best examples of the system, which overcame the disaster and recovered to its normal state. In numerical simulation to be described below some of the system parameters were evaluated using, for example, the real data for Fukushima prefecture. Namely, the total number of residents is evaluated as $10^6$, the area of the region treated as a certain administrative unit responsible for mitigating the aftermath is set about $10^4$~km$^2$, the mean distance between the neighboring cities in this region is 40--50~km, as a results, the number of cities that can be involved into recovering the affected region may be about 5--50.

\subsection{System under consideration}\label{sysconsid}

The system is modeled as a collection of cities connected with one another by a transport network. Initially in each city $i$ there is some amount of resources $Q_i$ depending on the population $N_i$ of residents. Under the normal conditions this amount of resources is excessive and substantially exceeds the minimal critical level $Q_{ci}$ required for its residents to survive during a certain period of time, $Q_i > Q_{ci}$. In order to clarify the introduction of the critical amount $Q_{ci}$ let us note the following. The phase of recovery process under consideration is characterized by a relatively short duration. In this case the demand of vital resources (quantity/time) is reduced to the critical amount  $Q_{ci}$ (quantity) evaluated from the expected duration of  short-term recovery implementation. Naturally for long-term processes like the long-term recovery phase the minimal critical amount of resources and the resource consumption should be considered individually. 

The critical amount of resources $Q_{ci}$  depends on the population $N_i$ of residents in a given city $i$; the larger the population, the higher the required level of resources $Q_{ci}$. One of the consequences of a large scale disaster is  increased demand for the vital resources in the affected cities. This is modeled as the essential increase in the corresponding value of $Q_{ci}$ and the opposite inequality $Q_i < Q_{ci}$ holds for the affected cities. Naturally the inequality
\begin{equation}\label{in:1}
\sum_i Q_i > \sum_i Q_{ci}
\end{equation}
must hold still after the disaster. Actually inequality (\ref{in:1}) is the mathematical implementation of the requirement that the given system is capable to survive as a whole during a certain length of time without external help.

To examine the dynamics of supply process two particular cases of the initial resource distribution will be modeled. The first one is a ``uniform'' system. In this case all the cities are supposed to be initially equal in the all parameters. The second one is a ``centralized'' system which comprises a collection of small cities (``satellites'') and ``centers''.  In the ``satellites"  population  is less than in the ``centers" and for them the equality $Q^{satelit}_i=Q^{satelit}_{ci}$ is assumed to hold at the initial stage. To make the systems comparable the total amount of resources and population
\begin{equation}\label{Eq.resource balance}
\sum_i Q^\text{centalized}_i = \sum_i Q^\text{uniform}_i
\end{equation}
\begin{equation}\label{Eq.residents balance}
\sum_i N^\text{centalized}_i = \sum_i N^\text{uniform}_i
\end{equation}
are supposed to be equal.

An example of the ``centralized'' system is illustrated schematically in Fig.~\ref{F1}. It depicts a collection of cities linked to one another with a transportation network. Its part affected by the disaster is shown as a shadowed region.

\subsection{Resource redistribution problem: general formulation}

To elucidate the basic features of the algorithm to be constructed, let us consider a characteristic example of the problem describing the resource redistribution during the short-term recovery accepting some simplifications without losing generality. Following the previous Section~\ref{sysconsid} the system at hand is a collection of cites $\mathbb{S}=\{i\}$ connected with one another via a road network whose origin-destination matrix $\mathbf{D}=\|d_{ij}\|$ specifies the travel time from city $i$ to city $j$. Before a disaster occurring, in each city $i$ there is a certain amount of vital resources $Q_{i}(0)$ equal or exceeding its critical level $Q_{ci}$ required under the normal conditions. As a result of disaster in the affected cites $\mathbb{A}$ the critical level $Q_{ci}$ immediately increases to $Q_{ci}^{\text{aff}}$ and becomes larger than the current amount of  vital resources, while; in the unaffected cities $\mathbb{U}$ the situation does not change. In other words, just after disaster at the initial time $t=0$ 
\begin{equation} \label{form:1}
\begin{split}
	Q_{i}(0)& \geq Q^{\phantom{\text{aff}}}_{ic} \qquad \text{for $i\in \mathbb{U}$}\,,\\
	Q_{i}(0)& < Q^\text{aff}_{ic} \qquad \text{for $i\in \mathbb{A}$}\,.
\end{split}
\end{equation}
During the short-term recovery the vital resources should be delivered to the affected cities from the unaffected ones and the resource redistribution process is terminated at a certain lenght of time $t=T_{f}$ when the state 
\begin{equation} \label{form:2}
\begin{split}
	Q_{i}(T_{f})& \geq  Q^{\phantom{\text{aff}}}_{ic} \qquad \text{for $i\in \mathbb{U}$}\,,\\
	Q_{i}(T_{f})& = Q^\text{aff}_{ic} \qquad \text{for $i\in \mathbb{A}$}\,.
\end{split}
\end{equation}
is achieved. The inequality being actually a rewritten form of \eqref{in:1}
\begin{equation} \label{form:3}
	\sum_{i\in\mathbb{S}}Q_{i}(0) > \sum_{i\in\mathbb{A}}Q^\text{aff}_{ic} + \sum_{i\in\mathbb{U}}Q_{ic} 
\end{equation}
is assumed to hold to make this redistribution feasible.

The resource transportation process is described as a collection of elementary events $\{e\}$ of carrying a certain fixed amount of the required commodities, the resource quantum $h$. Each elementary event is specified by four quantities
\begin{equation} \label{form:4}
	e = \{i,j,t,t'\} 
\end{equation}
meaning that a resource quantum $h$ is sent from city $i\in\mathbb{U}$ to city $j\in\mathbb{A}$ at a time moment $t$ and the destination gets the quantum time moment 
\begin{equation} \label{form:5}
	t' = t + d_{ij}\,. 
\end{equation}
Two factors are assumed to limit the resource transportation. The first one is the travel time between the cities. The second one is a certain time $\tau$ required for preparing a resource quantum to be transported with one vehicle from a warehouse located in a city. The value $\tau$ is considered to be the same for all the warehouses. The capacity $c_i$ of the warehouse at a given city $i$ is measured in the maximum number of vehicles that can be served simultaneously. The fixed value of $\tau$ enables us to regard the time moments of sending resource quanta as discrete variables, $t=t_n = n\tau$ where $n = 0,1,2,\ldots$  For the sake of simplicity the travel time $d_{ij}$  between any couple of cities $\{ij\}$ will be assumed to be also an integer number of time scales $\tau$, i.e.,  $d_{ij} = n_{ij}\tau$ and $n_{ij} = 1,2,3,\ldots$ As a consequence, the time $t'$ of resource quantum arrival is again an integer number of time unit $t'=n'\tau$ ($n' = 1,2,3,\ldots$). Implementation of the resource redistribution as a whole can be represented as a collection $\mathcal{P}=\{e\}$ of all these elementary events.

Generally speaking, any collection $\mathcal{P}= \{e\}$ of elementary events of currying resource quanta may be regarded as a virtual implementation of resource redistribution no matter how feasible or infeasible it is. To describe the analyzed problem let us introduce the universal set of these  elementary events
\begin{equation} \label{form:6}
	\mathcal{U}= \{e\},\ \text{where\ } i\in\mathbb{U}\,,\ j\in\mathbb{A}\,,\ \text{and $t=n\tau$ ($n = 0,1,2,\ldots$)} 
\end{equation}
The arrival time $t'$ can be also included into definition~\eqref{form:6} provided condition~\eqref{form:5} is taken into account. We call any subset $\mathcal{P}\subset\mathcal{U}$ a virtual plan of the resource redistribution implementation. 

For a virtual plan $\mathcal{P}$ to become a realistic one several conditions should be fulfilled. To write these requirements we introduce the following quantities determined for each virtual plan $\mathcal{P}$.  They are the number of vehicles leaving a given warehouse $i$ at a time moment $t$
\begin{equation} \label{form:7}
	I^-_{it}\{\mathcal{P}\}= \sum_{e\in\mathcal{P}} \delta_{ii_e}\delta_{tt_e}\,,\qquad i\in\mathbb{U}\,,
\end{equation}
where $\delta_{ij}$ is Kronecker's delta ($\delta_{ii} =1$ and $\delta_{ij} = 0$ for $j\neq i$) and the subscript $e$ has been added to the components of the event $e=\{i_e,j_e,t_e,t'_e\}$ to distinguish them from the corresponding indices of cities at hand and the analyzed time moments. The second collection of quantities is the number of vehicles arriving at an affected city $j$ at a time moment $t'$,    
\begin{equation} \label{form:8}
	I^+_{jt'}\{\mathcal{P}\}= \sum_{e\in\mathcal{P}} \delta_{jj_e}\delta_{t't'_e}\,,\qquad j\in\mathbb{A}\,.
\end{equation}
In these terms the conditions imposed on any realistic plan $\mathcal{P}$ are as follows. First, due to the \textit{limit capacity} of warehouses, for each city we write
\begin{align} 
\label{form:9}
	I^-_{it}\{\mathcal{P}\} &\leq c_i &&\qquad\text{for\ } \forall\, t\text{\ and\ } \forall i\in\mathbb{U} \,.\\
%
\intertext{Second, \textit{the resource conservation} in the undamaged cities is represented as}
\label{form:10}
	h\sum_t I^-_{it}\{\mathcal{P}\} &\leq Q_i(0) - Q_{ci} &&\qquad\text{for\ } \forall i\in\mathbb{U} \,.\\
\intertext{Third, \textit{the relevancy of the resource delivary} to the affected cities becomes}
\label{form:11}
	h\sum_{t'} I^+_{jt'}\{\mathcal{P}\} & \stackrel{h}{{} ={}} Q^\text{aff}_{cj}-Q_j(0) &&\qquad\text{for\ } \forall j\in\mathbb{A} \,.
	\end{align}
The symbol `$\stackrel{h}{=}$' in the above expression means the equality within the accuracy of one resource quantum $h$; if all the quantities of resource amount are some integer numbers of quanta this symbol is just the rigorous equality. In what follows for the sake of simplicity we will not distinguish between them.  

The dynamics of the resource redistribution is governed by the equations	
\begin{align}
\label{form:add1a} 
	Q_{jt'+\tau} & = Q_{jt'}+ I^+_{jt'}  &&\qquad\text{for\ } \forall\, t'\text{\ and\ } \forall j\in\mathbb{A} \,,\\
\label{form:add1b} 
	Q_{it+\tau} & = Q_{it} - I^-_{it}  &&\qquad\text{for\ } \forall\, t\phantom{'}\text{\ and\ } \forall i\in\mathbb{U} \,,\\
\intertext{subject to the initial condition} 
\label{form:add2}
	Q_{kt}|_{t=0} &= Q_{k0} && \qquad\text{for\ } \forall k\in\mathbb{S} \,. 
\end{align} 
The time moment $T_j=T_{j}\{\mathcal{P}\}$ when the condition 
\begin{equation} \label{form:add2a}
Q_{jt'}|_{t'=T_j}< Q^\text{aff}_{cj}\quad\text{and}\quad Q_{jt'}|_{t'=T_j+\tau}= Q^\text{aff}_{cj}
\end{equation}
is fulfilled will be refereed to as the time of completing the resource redistribution with respect to city $j\in\mathbb{A}$.

\subsubsection*{Standards of resource supply}

Now let us discuss the principles making a given plan $\mathcal{P}$ acceptable for implementation. There are two factors different in nature impacting on the resource supply within  short-term recovery. 

The first one is categorized as the \textit{process efficiency} and reduced to the requirement of minimizing the duration of  resource redistribution during  short-term recovery (see Introduction). If we had confined the constrains imposed on this process to conditions~\eqref{form:9}--\eqref{form:11} than the given requirement would give rise to 
\begin{equation} \label{form:last1}
 \underset{\mathcal{P}\subset\mathcal{U}}{\text{minimize}} 
 \left\{\max_{j\in\mathbb{A}}\Big[T_j\{\mathcal{P}\}\Big]\right\}\quad
 \text{subject to \eqref{form:9}--\eqref{form:11}.}
\end{equation}
It is a conventional optimization problem admitting also some generalization to account for the contribution of all the quantities $\big\{T_{j}\{\mathcal{P}\}\big\}$. However, the short-term recovery is a rescue operation directly aimed at human life saving, which imposes  another requirement irreducible to formal optimization of some functiona on the resource redistributionl.  
  
In rescue operations under various conditions humanitarian factors play crucial roles. We keep in mind one of them that can be categorized as the \textit{ethic or morality criterion} in the priority choice during the operations. Let us appeal, e.g., to a historical example. Baron Dominique Jean Larrey, surgeon-in-chief to Napoleon's Imperial Guard, articulated one of the first triage rule in 1792: ``Those who are dangerously wounded should receive the first attention, without regard to rank or distinction. They who are injured in a less degree may wait until their brethren in arms, who are badly mutilated, have been operated on and dressed, otherwise the latter would not survive many hours; rarely, until the succeeding day" \cite{Napaleon}. This principle has given rise to the concept of triage in hospitals worldwide (for an introduction and the relevant literature see, e.g, \cite{iserson2007triage1,iserson2007triage2} as well as \cite{wiki:Triage}).

In order to elucidate how the humanitarian factors can be incorporated into a minimizing problem like \eqref{form:last1} we note that, first, rather arbitrary conditions can be added to \eqref{form:9}--\eqref{form:11}. The latter three constraints just make a trial plan $\mathcal{P}$ feasible so newly added conditions simply should not be in conflict with them. Second, any plan generated by model~\eqref{form:last1} or a similar one is strictly optimal provided all the information is known beforehand. However, in the reality it is hardly feasible, moreover, the situation in the case of a large scale disaster can change unpredictably. Appealing to the experience accumulated by the human society for many centuries we may expect that some rules how to cope with such situations were found and converted into moral and ethic norms (a discussion of this point of view can be found, e.g, in \cite{tannert2007ethics}). It allows us to regard the triage principle as an additional condition that should be added to problem~\eqref{form:last1}. It makes the corresponding plan not strictly optimal as a mathematical solution of \eqref{form:last1} but the result should be more attractive for human society and can be more adequate for actions under uncertainty. Naturally, the latter factor is worthy of individual consideration. In the present paper we just confine our consideration to the optimizing problem~\eqref{form:last1} subject in addition to the \textit{triage principle}.


It is presumed that at any moment of time $t$ it is possible to single out one or several cities $\mathbb{E}_t\subset\mathbb{A}$ where the emergency level of their residents surviving is currently \textit{highest} among the other damaged cities. A certain priority function 
\begin{equation} \label{form:last2}
S_j(t) = S\big(Q_{jt},Q^\text{aff}_{cj},N_j\big).
\end{equation}
given beforehand is used to quantify this emergency level,
\begin{equation} \label{form:last3}
S_{j}(t) = \max_{k\in\mathbb{A}}\{S_k(t)\}\qquad\text{for\ }\forall j\in \mathbb{E}_t\,.
\end{equation}
For the sake of simplicity the state of each city is evaluated using the same function whose list of arguments includes the current amount of the required resources $Q_{jt}$, their critical level $Q^\text{aff}_{cj}$, and the population $N_j$, other possible arguments such as the type of commodity are not shown explicitly.  

The triage principle is implemented via the requirement that at any moment of time $t$ the required resources, at first, be directed to the cities $\mathbb{E}_t$ with the worst situation. The vital resources may be sent to other cities if it does not interfere with the previous action. So in the general case the triage principle does not specifies exactly the system dynamics, it imposes  it some restrictions within which the system can be governed by other mechanisms, in the given case, it is the minimization of delivery time.   

\begin{figure}
\begin{center}
\includegraphics[width=0.5\columnwidth]{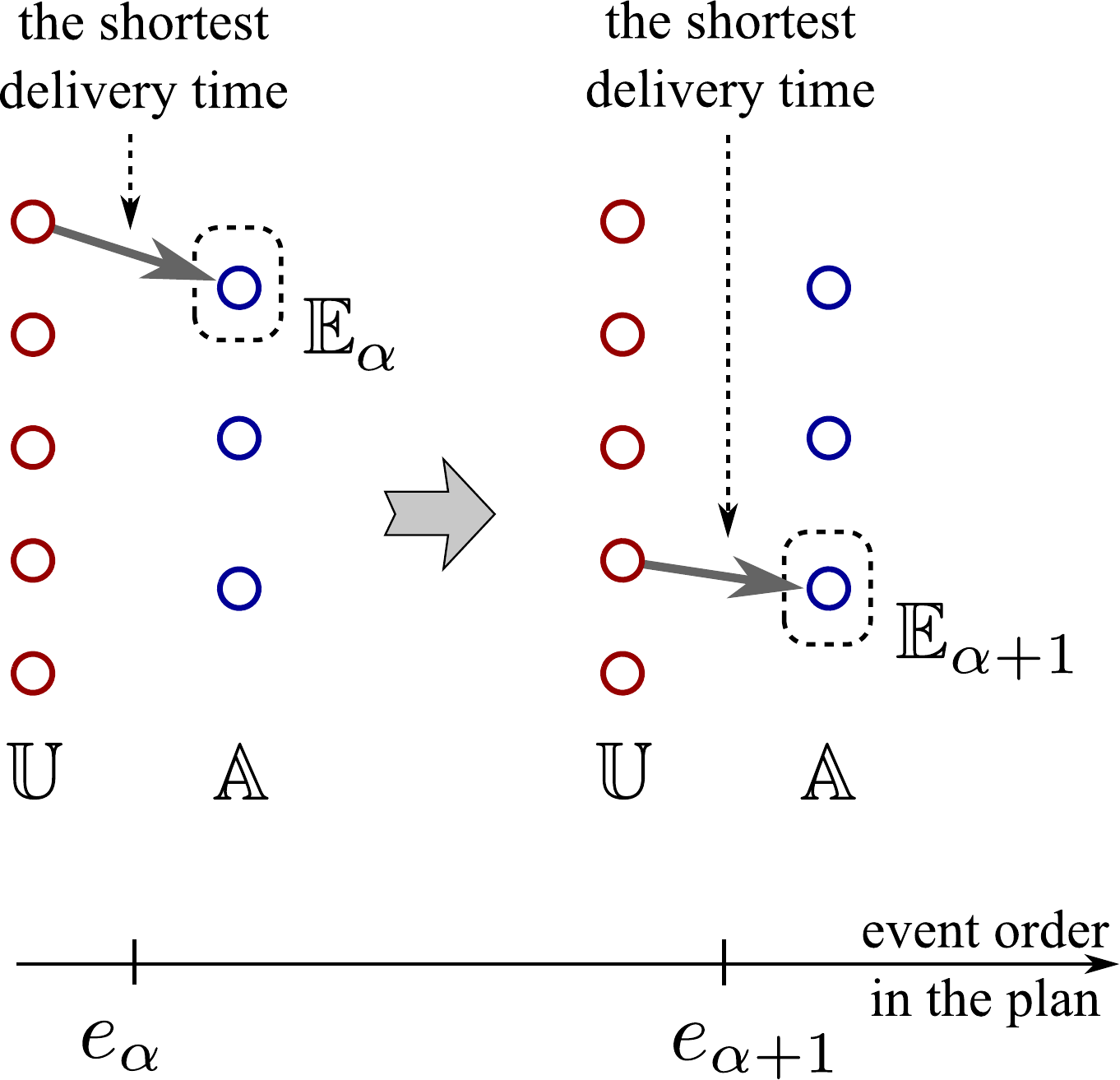}
\end{center}
\caption{A schematic illustration of the ``optimal'' plan formation governed by the triage principle and minimizing the resource delivery time.}
\label{F:last1}
\end{figure}

How this combination of the triage principle and the minimization of the resource delivery time is used in constructing the ``optimal'' plan of the resource redistribution is described in detail in the next Section. Here we discuss its general aspects illustrated in Fig.~\ref{F:last1}. The point is ordering the elementary events $\{e_\alpha\}=\mathcal{P}$ in the analyzed transportation plan $\mathcal{P}$ in a way matching the triage principle. It means that the city $j\in\mathbb{A}$ to be directed a resource quantum may be chosen only from the current priority set $\mathbb{E}_\alpha$. Within this restriction the specific choice of the elementary event $e$ is subjected to the minimality of the delivery time provided the warehouse to be chosen is physically able to send the required resource quantum. In the model at hand a vehicle staring its motion unconditionally gets the destination. It is possible to consider that the action of sending a quantum can change immediately the priority set of cities at the next step
\[
\mathbb{E}_{\alpha}\stackrel{\substack{\text{resource}\\ \text{sending}}}{\Longrightarrow} \mathbb{E}_{\alpha+1}   \,.   
\]
At the next $\alpha+1$ step the described actions are repeated again. It should be noted that in constructing a plan in this way the resources are sent virtually and the order of the elementary events $\{e_\alpha\}$ in the plan $\mathcal{P}$ does not mandatory conserve the real time direction. The triage principle requires only the worst case to be the first in decision making. The other cases may be proceeded earlier in the real time if it does not interfere the first case.

\section{Resource redistribution algorithm}

This section presents the logic of resource redistribution and its realization algorithm. We note that there is an essential difference between the problem under discussion and problems of classical logistics; in our case the network is dynamical. All the resources are located in some warehouses in or near the cities and their capacities are limited with respect to the amount of resources as well as the operation ability (limited number of loading vehicles). In particular, when the first group of parcels in one of the warehouses are sent, it takes a time to prepare another one for sending. During this interval the resources in the given warehouse are not accessible for all the other cities and this warehouse became temporally ``cut off" from the network. Such behavior of the system endows a nonlinear process. To take this effect into account the algorithm uses the time distance between cities instead of geographic one and their specific values depend on the intensity of supply flow.

At the initial step all the cities that are accessible provide the information about their state, namely, the available amount of resources $Q_i$, the minimal critical amount $Q_{ci}$ required for their individual surviving, and the population $N_i$. The characteristics of the transportation  are assumed to be also given, it is the matrix $\mathbf{D}=\|d_{ij}\|$ whose element, e.g., $d_{ij}$ specifies the minimal time distance between city $i$ and $j$. To describe the states of cities let us introduce the value
\begin{equation}\label{eq:theta}
	\theta_i = \frac{Q_i - Q_{ci}}{Q_{ci}}\,.
\end{equation}
If the information about a given city $i$ is not available, then the corresponding value is set equal to zero, $\theta_{i}=0$. When $\theta_i < 0$ its magnitude quantifies the lack of vital resources in relative units. The quantity $S_i = \theta_i N_i$, or more strictly its absolute value is actually the number of people being under the level of surviving. It will be used in specifying the priority of the cities in the resource redistribution queue. The minimal value of $S$ corresponds to the maximal number of residents which are not supplied with vital resources and it endows us to mark that city as  most ``dangerously wounded''.   

Because the main goal of resource redistribution just after the disaster is mitigation of consequences and minimization of the amount of victims, Table~\ref{T1} determines the priority of the resource redistribution.
\begin{table}[h]
\caption{The order of cities according to the resource redistribution priority. Here $M$ is the total number of cities in the given system.}\label{T1}
\begin{center}
\begin{tabular}{|c||c|c|c|c|c|}
\hline
$S_p$ & $S_1$ & $S_2$ & $\ldots$ & $S_{M-1}$ & $S_{M}$ \\
\hline
$p$ &   $1$ &  $2$ &  $\ldots$ &    $M-1$ &  $M$  \\
\hline 
\end{tabular}
\end{center}
\end{table}
\noindent
The order used in Table~\ref{T1} matches the inequality
\begin{equation}\label{eq:order}
S_1 \leq S_2 \leq \ldots\leq S_{M-1}\leq S_{M}
\end{equation}
and $i_1$, $i_2$, $\ldots$, are the corresponding indexes of these cities. 

In order to describe resource redistribution dynamics, let us introduce the following quantities. First, it is a certain quantum $h$ of resources that can be directed from a city to another one. The second quantity is the time $\Delta t$ required for this quantum to be assembled for transportation. The third one  $c_i$ is the capacity of a given city $i$ specifying the maximal number of quanta which can be assembled during the time $\Delta t$. Introduction of these quantities implies the realization of resource redistribution mainly via fast loading vehicles, for instance, tracks. In this case $h$ is the volume of resources transported by the typical vehicle individually, $\Delta t$ is the time necessary to load this vehicle, and $c_i$ is determined by the number of loading places and the capacity of loading facilities. 

The algorithm to be described below creates a complete plan of resource redistribution depending explicitly on the initial post-disaster system state. Namely, at the first step Table~\ref{T1} is formed using the initial data. The city $i_1$ is selected as the city with the wost situation. Then we choose a city $i_k$ such that
\begin{equation}\label{eq:dsearch}
d_{i_k i_1 } = \min_j d_{j i_1}\quad \text{among} \quad Q_j - h \geq Q_{cj}\,.
\end{equation}
Then the prepared quantum is \textit{virtually} transported to city $i_1$ from city $i_k$ . It gives rise to the transformations
\begin{equation}\label{eq:transf}
\begin{split}
Q_{i_1} &\rightarrow Q_{i_1} + h\,,\\ 
Q_{i_k} &\rightarrow Q_{i_k} - h\,, \\ 
c_{i_k} & \rightarrow c_{i_k} - 1\,.
\end{split}
\end{equation}
The information about the given action is saved as a report of its virtual realization and comprises: ``city $i_k$ sent one quantum to city $i_1$ at time $t_{\text{dep},i_k}$, the quantum is received at time $t_\text{arr}$''. 
Initially for all the cities involved in the resource redistribution we set $t^\text{init}_{\text{dep},i_k}=0$. The further modification of these values will be explained below, see Eq.~\eqref{eq:decrD}.
In the developed algorithm the time moments $\{t_\text{arr}\}$ are specified via the expression
\begin{equation}\label{eq:tdeptarr}
t_\text{arr} =  d_{i_k i_1}\,. 
\end{equation}
It should be noted that formula~\eqref{eq:tdeptarr} obviously holds at the initial steps when $t^\text{init}_{\text{dep},i_k}=0$ and the original element of the matrix $\mathbf{D}$ enters it. Its use in the general case will be justified by the renormalization of the matrix  $\mathbf{D}$ (see Eq.~\eqref{eq:decrD}) taking into account the delay in sending the resource quanta caused by the city limit capacity leading to nonlinear effects in the resource redistribution.

At the next step this procedure is reproduced again. Table~\ref{T1} is reconstructed, the logic of choosing the interacting cities is repeated with saving the relevant report. 

Since the maximal number of quanta that can be sent from a given city simultaneously is finite, there exist a situation where $c_j$ takes a zero value due to transformations~\eqref{eq:transf}. This effect is taken into account by renormalization of the matrix $\mathbf{D}$, which is a time distance matrix. Namely, when $c_j = 0$ we restore the initial value of $c_j$ and for all $i$
\begin{equation}\label{eq:decrD}
\begin{split}
d_{ji} &\rightarrow  d_{ji} + \Delta t\,,\\ 
t_{\text{dep},j} &\rightarrow t_{\text{dep},j}+ \Delta t\,.
\end{split}
\end{equation}
This procedure is terminated when at the next step 
\begin{equation}
\forall i: \quad S_i > 0\,.
\end{equation}
As a result, this algorithm generates the collection of reports which enables us to create a semi-optimal plan of resource redistribution for all the cities and the real process of resource redistribution is initiated.

According to this plan the cities start sending the real resources. If at a certain moment of time $T_1$ new information about the system state is received the procedure of plan construction is repeated. This reconstruction takes into account two factors. First, it is the new data about the city damage $\{Q_{ci}\}$. The second factor is the current pattern of resource allocation in the system, $\{Q_{i}(T_1)\}$. It is determined by the implementation of the previous plan of sending the resource quanta before the moment $T_1$. Within the frameworks of the developed algorithm this construction is possible, because it is based on the collection of reports like ``city $i_k$ sent one quantum to city $i_1$ at time $t_\text{dep}$, the quantum is received at time $t_\text{arr}$'' and at any time moment it is possible to figure out how many resource quanta have been sent, got the destination, and are in the transportation process. Thereby, the new plan replaces the previous one from the time moment $T_1$ and continues governing the further resource redistribution. In the case of a new update event such reconstruction is repeated again.

\section{Numerical Simulation}

Two principally different situations were studied numerically. The first one is the case when the whole information about the state of the system is available initially. Under this condition the resource redistribution process is studied depending on system parameters such as the damage level, the number of cities, the type of the initial resource allocation, and the city capacity. In the second situation the information about the system state is updated gradually during the process, which enables us to analyze the effect on uncertainty on the short term recovery.

\begin{figure}
\begin{center}
\includegraphics[width=0.4\columnwidth]{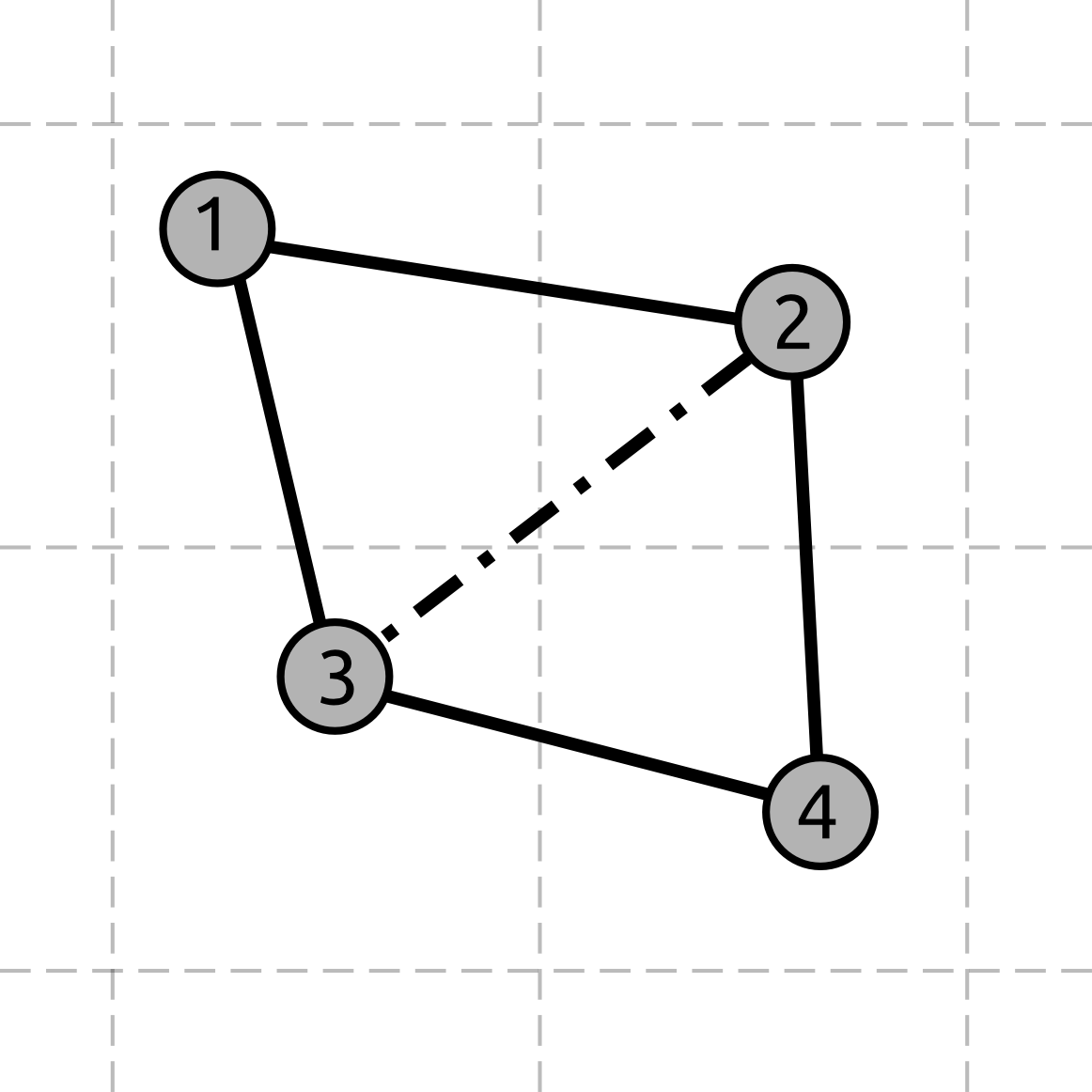}
\end{center}
\caption{Example of the city arrangement and the corresponding transportation network.}
\label{F:angle}
\end{figure}

\subsection{Details of modeling: case with available information}\label{sec:det_availinf}

The purpose of the present section is to illustrate the features of the analyzed resource redistribution process. Keeping in mind the administrative units noted in Section~\ref{modback}, the following systems were studied numerically based on the developed model. Each of them is assumed to comprise 20 cities regarded as basic entities connected with one another by a transport network and the total population of these cities is set $P=2\times 10^6$. Two types of systems, ``uniform'' and ``centralized'' were analyzed separately. For specific purposes some system parameters, namely, the number of cities and the total population were changed.  The amount of resources were measured by the unit of resource quantum $h$, so we set $h=1$. To be specific the volume of one quantum is assumed to afford 100 residents with some additional extra volume (60~\%) under the normal conditions. So the integral amount of resources initially allocated in the system is 
\begin{equation*}
\sum_i Q_i = \frac{P}{100}\,.
\end{equation*}
The mean time distance between the cities was varied from 40 to 120 minutes and the time $\Delta t$ necessary to prepare one resource quantum was set 5--15 minutes.

The transportation network was constructed in the following way. The region occupied by the given system is considered to be of a rectangular form and divided into $20$ (the number of cities) equal rectangles. Each rectangle contains one city placed randomly within it. At the first step the connections between the cities located in the neighboring rectangles are formed as illustrated in Fig.~\ref{F:angle}. For any arrangement of these four cities the "vertical" and "horizontal" connections are formed. A diagonal connection, for example, the connection 2-3 is formed if both of the opposite angles are less than $90^\text{o}$: $\angle{213}$ and $\angle{243}$ in Fig.~\ref{F:angle}. In this way we construct the matrix \textbf{D} of minimal time distances between the neighboring cities. The relationship between spatial and temporal scales was determined assuming the average speed of transporting vehicles equals 60 km/h. At the next step using Warshall's algorithm (see, e.g., \cite{discmathstr_kbr6}) we complete the matrix  \textbf{D} of the minimal time distances between any pair of cities. In the case of affected cities some of
the connections were cut up, however, without losing the graph connectivity.

\begin{figure}
\begin{center}
\includegraphics[width=0.6\columnwidth]{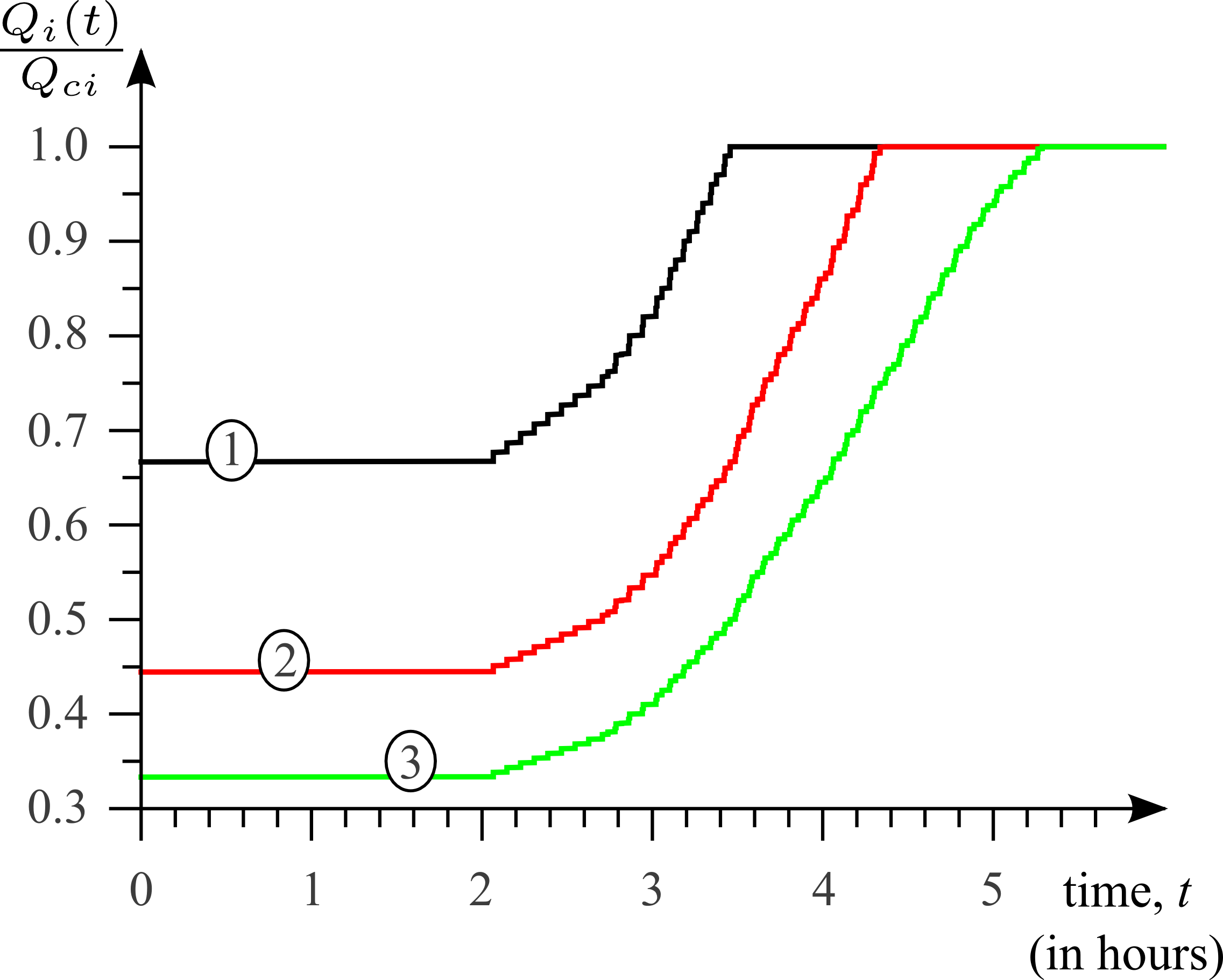}
\end{center}
\caption{The recovery dynamics of an affected city $i$. Curves 1, 2, 3 demonstrate the recovery dynamics when the degree of damage $a$ is equal to 1.5, 2.25, 3, respectively. All the other system parameters as well as the system topology were the same.}
\label{F:D123}
\end{figure}

\subsection{``Uniform'' system}

First, let us consider the results of numerical simulation for the ``uniform'' system. Three neighboring cities located at one of the rectangular corners were supposed to be damaged. Figure~\ref{F:D123} illustrates the dynamics of supplying an affected city $i$ according to the plan generated by the developed algorithm. Three curves represents the recovery dynamics of the affected city $i$ for three different degrees of damage, $a = \dfrac{Q^\text{affected}_{ci}}{Q_i(0)} = 1.5,\ 2.25,\ 3$, and it explains the difference in the initial values of $\dfrac{Q_i(0)}{Q_{ci}}$ for the curves. 

As far as the general shape of these curves is concerned, it is similar to the classical resilience triangle (see, e.g.,  \cite{bruneau2003framework}) as should be expected according the modern concept of the recovery processes. Namely, the initial horizontal fragment ends when the first resource quantum reaches the given city, the intermediate fragment exhibits the recovery to the minimal operating standards followed by the saturation meaning the finishing of the short-term recovery.

\begin{figure}
\begin{center}
\includegraphics[width=\columnwidth]{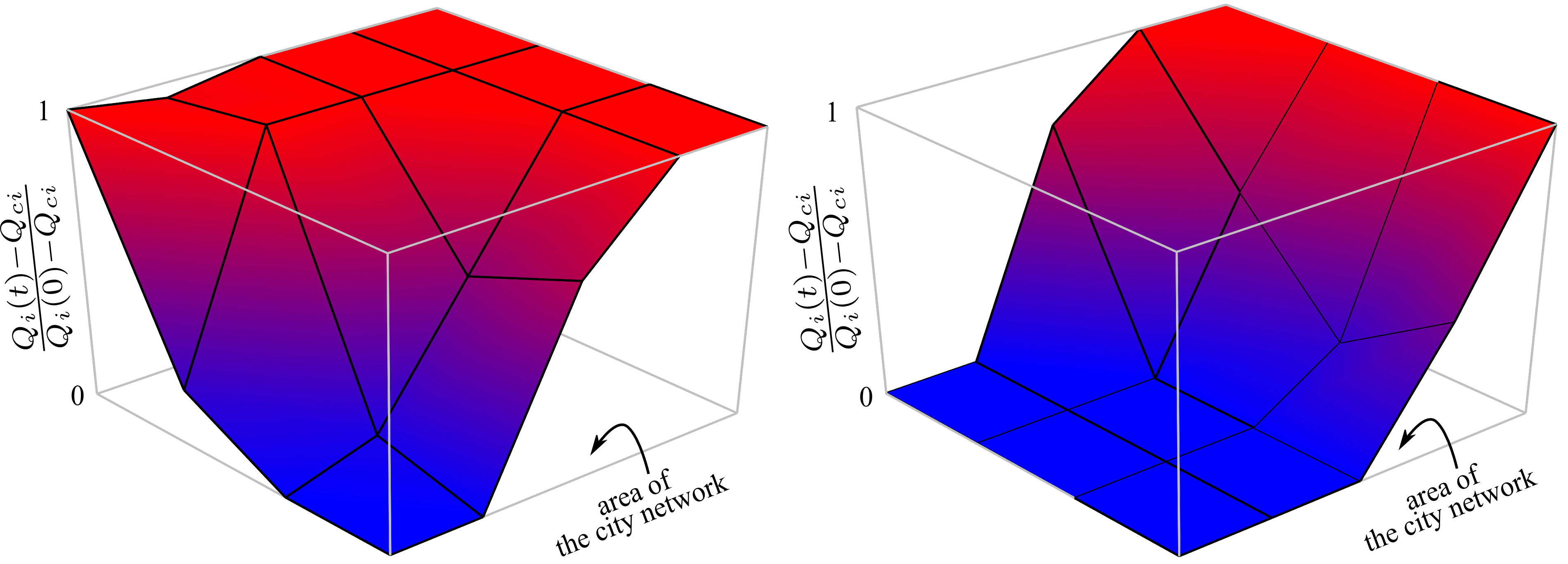}
\end{center}
\caption{The spatial pattern of the extra volume of resources distributed in the system after the recovery process has finished for two different degrees of damage, $a = 1.5$ (left) and $a = 3$ (right). As clearly shown, in the second case the number of cities involved in resource redistribution is considerably more than in the first case.}
\label{F:PatD123}
\end{figure}

The present result demonstrates a significant influence of the cooperative effects on the recovery dynamics. In fact, let us compare Case~1 and Case~3 (Fig.~\ref{F:D123}). The number of quanta required to recover city~$i$ inCase~3 is four times bigger than that of Case~1. However, the total duration of the redistribution process increases by less than twice. It is because that the greater damage is caused by the disaster, the more cities are involved in the recovery process. Figure~\ref{F:PatD123} justifies this conclusion depicting the spatial pattern of the extra volume of resources distributed in the system after the recovery process has finished.

\begin{figure}
\begin{center}
\includegraphics[width=0.55\columnwidth]{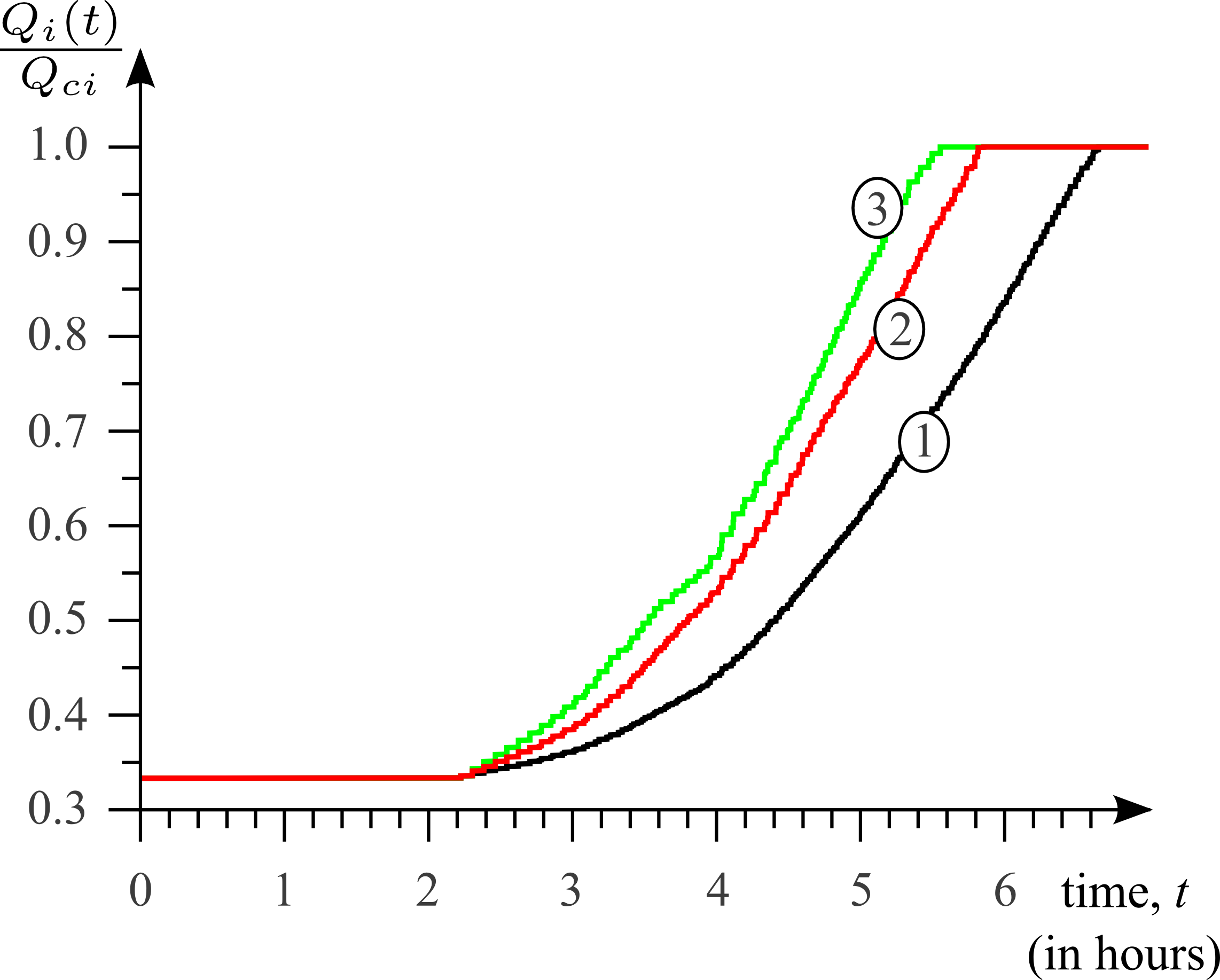}
\end{center}
\caption{The recovery dynamics of an affected city $i$. Curves 1, 2, 3 demonstrate the recovery dynamics when the capacity of cities  $c$ is equal to 15, 30, 45, respectively. All the other system parameters as well as the system topology were the same.}
\label{F:Q(t)C123}
\end{figure}

\begin{figure}
\begin{center}
\includegraphics[width=1.0\columnwidth]{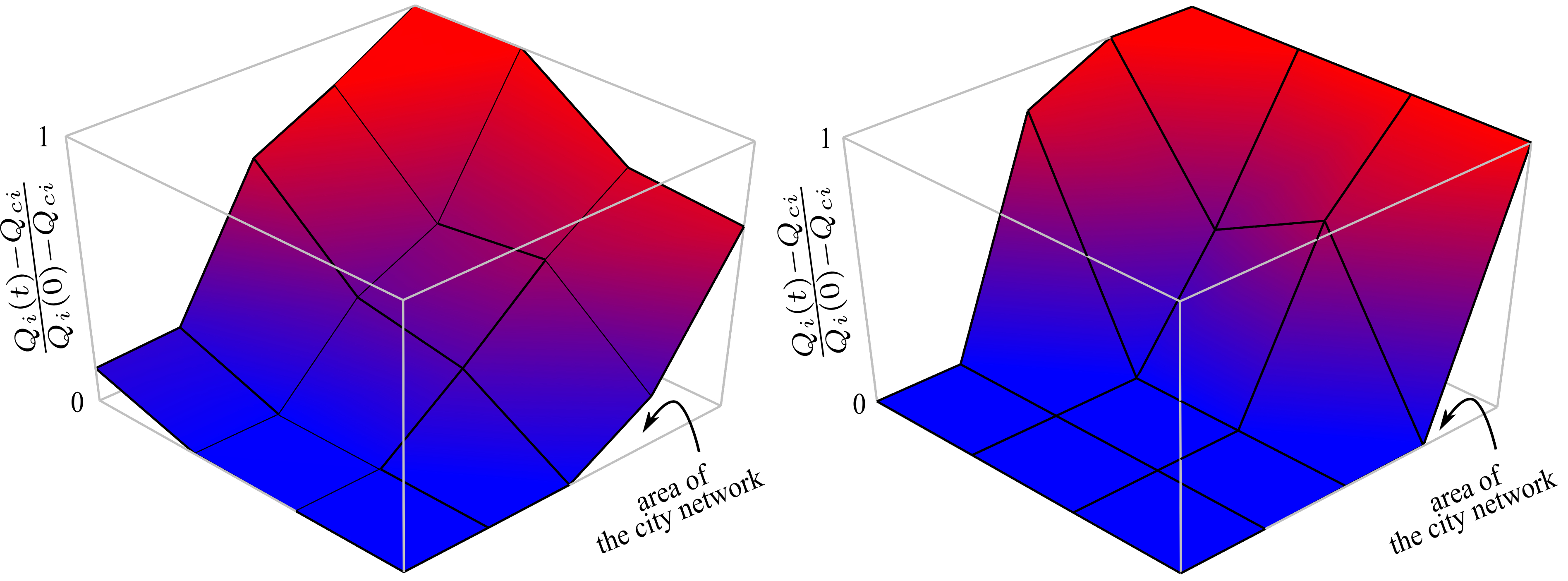}
\end{center}
\caption{The spatial pattern of the extra volume of resources distributed in the system after the recovery process has finished for two different values of the city capacity, $c = 15$ (left) and $c = 45$ (right). As clearly shown, in the second case the resource redistribution process is more local than in the first case.}
\label{F:PatC1C3}
\end{figure}

The next result is presented in Fig.~\ref{F:Q(t)C123} depicting the recovery dynamics of the affected city $i$ depending on the city capacity $c_i$ for a fixed damage degree, $a=3$. The capacity $c_i$ was changed from 15 to 45 for all the cities, which means that the number of quanta the cities are able to send per unit time was increased by three times in simulation. Nevertheless, the duration of recovery process changed only 1.5 hours (less than 30\% in relative units). It is also explained by the cooperative effects in the resource redistribution process, which is directly demonstrated in Fig.~\ref{F:PatC1C3}.

As shown in Fig.~\ref{F:PatD123}, both the quantities, the degree of damage $a$, and the city capacity $c_i$, affect the number of cities involved in the resource redistribution. One should, however, distinguish their effects. The degree of damage is not controllable parameter but a disaster characteristic, while the city capacity is a controllable technical parameter.

Some additional details of the effect caused by the city capacity are illustrated in Fig.~\ref{F:PatC1C3}. It depicts the spatial pattern of the extra volume of resources distributed in the system after the recovery process has finished for two different values of  the city capacity for the all cities, $c_i = 15$ (left) and $c_i = 45$ (right). We can see that in the case of a smaller capacity the number of cities involved in the redistribution process is more than in the case of a larger capacity. At the same time, the number of cities that send out all their extra volume of resources is more for the larger capacity. On one hand, therefore, the larger the capacity, the smaller the region comprising the cities involved in the recovery process, i.e., the higher the locality of this process. On the other hand, the smaller the capacity, the less the number of cities being in close to the minimal operation standards. Thereby the choice (if possible) of various values of $c_i$ can be determined for specific reasons.

\begin{figure}
\begin{center}
\includegraphics[width=1.0\columnwidth]{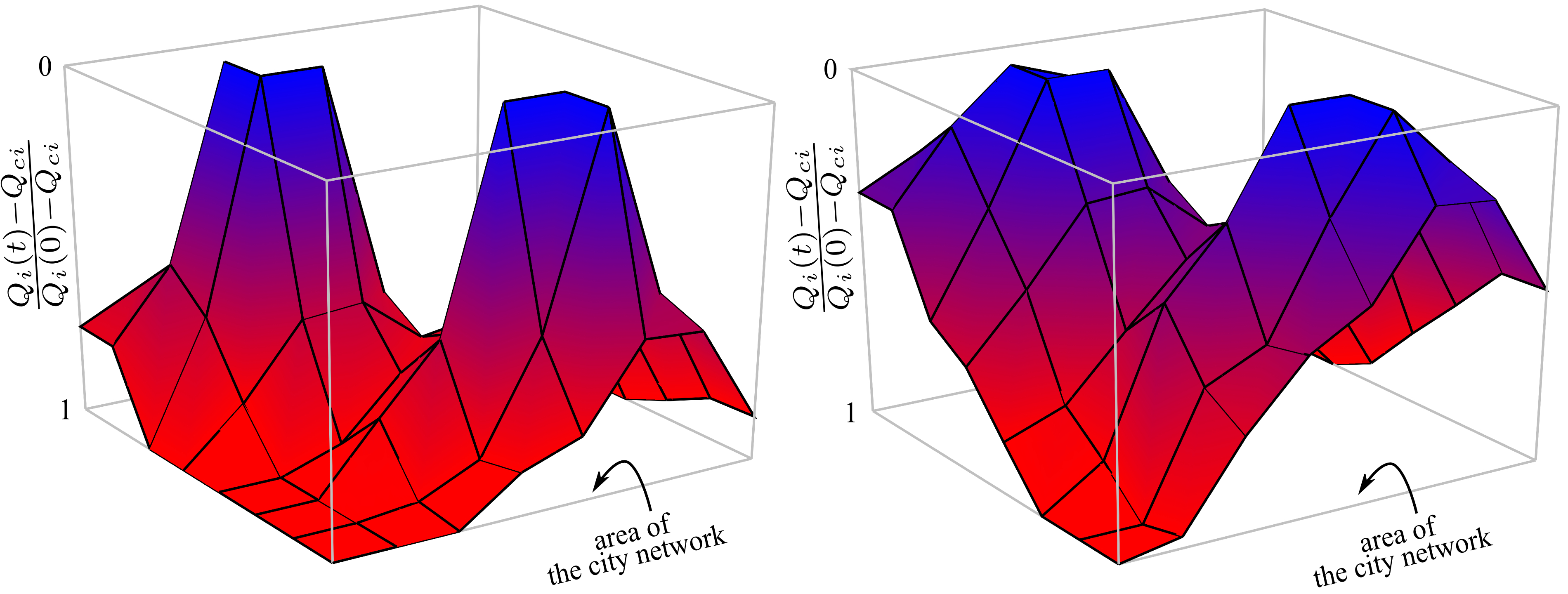}
\end{center}
\caption{The inverted spatial pattern of the extra volume of resources distributed in the system after the recovery process has finished when two separate regions interact directly. The left plot represents the spatial pattern for a relatively low degree of damage ($a = 1.5$); the right plot for a high degree of damage ($a = 3$).}
\label{F:Pat2AP}
\end{figure}

The effect of locality becomes more pronounced in the case where there are several separate groups of affected cities. To illustrate this we considered  a system with the increased number of cities from 20 to 64 and assumed that the affected cities belong to two groups located in the opposite sides of the system region. The population and the total amount of resources were increased proportionally. Figure~\ref{F:Pat2AP} exhibits the spatial pattern of the extra volume of resources distributed in the system after the recovery process has finished. The left plot presents this patten for a relatively low degree of damage ($a = 1.5$). The right plot shows it for a high degree of damage ($a=3$). As shown here, in the former case we can identify two subsystems that do not interfere with each other in resource redistribution. As the degree of damage grows, the redistribution process drives these subsystems to cooperate and operate as a whole. The latter case exemplifies this effect. The cities located in the middle of the system region became involved in the redistribution of resources for the both affected groups of cities. It explains the saddle-shaped surface shown in the bottom plot. We call this effect  ``interference''.

\begin{figure}
\begin{center}
\includegraphics[width=0.55\columnwidth]{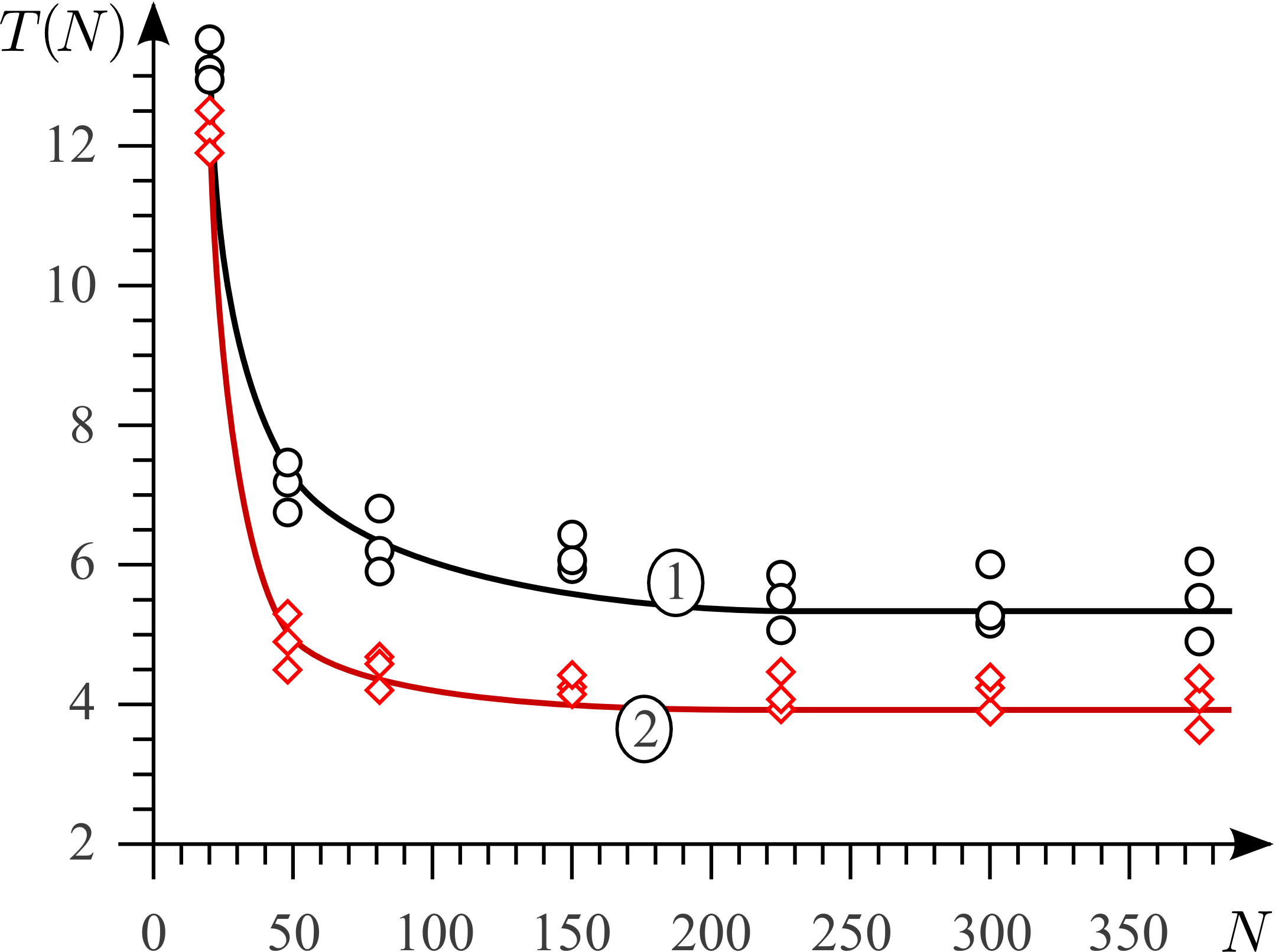}
\end{center}
\caption{The recovery duration $T(N)$  vs the number $N$ of cities that can be involved in the redistribution process. Curve~1 exemplifies the dependence for the case where the damaged cities are located in the corner, Curve~2 in the center of the system region. The dots represent the simulation results, and the curves are guides for the readers.}
\label{F:T(N)}
\end{figure}

The analyzed model uses a notion of administrative unit as an isolated system of cities individually responsible for the short-term recovery. Naturally, a cooperation of several administrative units can shorten the duration of this process. In order to study when such cooperation is efficient, we simulated the resource redistribution varying the number of cities (form 20 to 400) that can be involved in the process in principle. The result is presented in Fig.~\ref{F:T(N)} showing the dependence of the duration of redistribution process on the number of cities to be involved. The capacity of cities was set $15$, the time necessary to prepare one quantum of resources was increased by three times and set $0.25$ hour, and the degree of damage was set four. Curve~1 exemplifies this dependence for the case where the damaged cities are located in the corner, and Curve~2 in the center of the system region. For the given values of the system parameters the duration of the short-recovery exhibits fast drop within a interval from $20$ to $80$. It is explained by the fact that for this size of system all the cities are involved in the resource redistribution. When the system size exceeds some value around 100 cites, cities not participating in the process increase. When the distance between a given city and the affected region is far enough, it is more efficient to wait until a new quanta will be prepared in the neighboring cities than to request resources from the distance. It is responsible for the saturation in the dependence of the process duration vs the number of cities (Fig.\ref{F:T(N)}).

\subsection{``Centralized'' system}

Now let us consider the other type of system, i.e., the ``centralized'' one. In some sense it is the opposite type of the city network topology. Namely, we assumed that there are four big central cities of the equal size surrounded by 16 ``satellites'' (small cities), and 40\% of the system population are the residents of these centers. To compare the recovery dynamics for the ``centralized'' and ``uniform'' systems, the amount of critical resources was scaled with the population of residents such that the ratio $\dfrac{Q_{ci}}{N_i}$ is to be the same for the both types of systems.   
The amounts of resources in the ``satellites'' were set equal to their critical values $Q^\text{satellite}_i = Q^\text{satellite}_{ci}$, and all the ``extra'' resources of the system were concentrated in the centers such that $Q^\text{center}_i \gg Q^\text{center}_{ci}$. The total amount of resources and the system population were also equal for the ``centralized'' and ``uniform'' systems, i.e., Eqs.~\eqref{Eq.resource balance} and \eqref{Eq.residents balance}. The capacity $c_i$ of the centers was chosen twice as large as the city capacities for the ``uniform'' system, and the degree of damage was set $a = 2$. Besides, in order to smooth the discretization effects in the resource redistribution in the case under consideration we used the decreased volume of resource quantum, $h=0.2$, assuming it affords 20 residents.  

\begin{figure}
\begin{center}
\includegraphics[width=0.55\columnwidth]{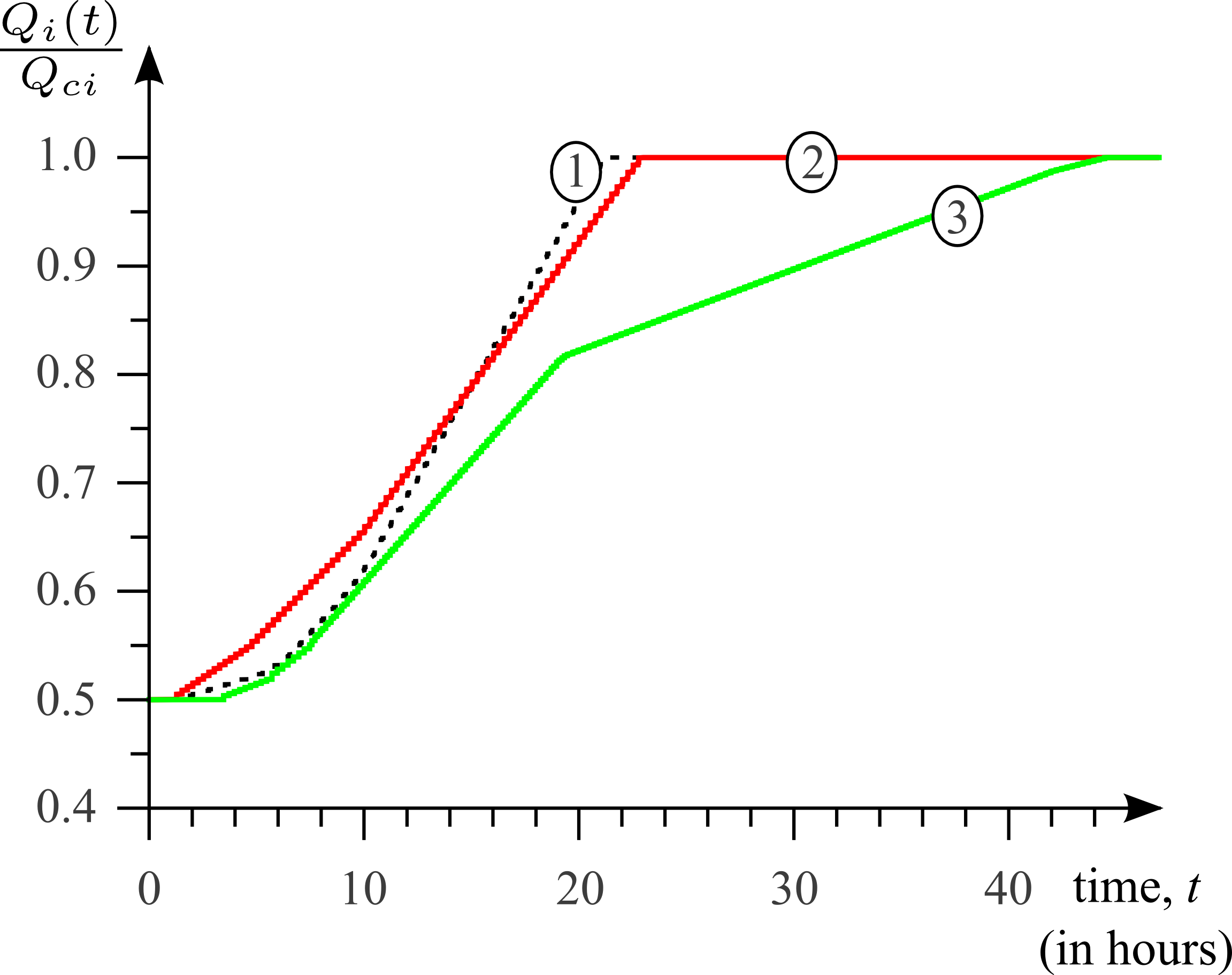}
\end{center}
\caption{The recovery dynamics of an affected city $i$ in three cases. Curve~1 demonstrates the recovery process in the ``uniform'' system, Curve~2 illustrates it when only ``satellites'' in the ``centralized'' system are damaged, and Curve~3 depicts the recovery dynamics of a damaged center.}
\label{F:Q(t)uc}
\end{figure}

Figure~\ref{F:Q(t)uc} compares the dynamics of the affected city $i$ in three cases. Curve~1 (dotted) shows the recovery dynamics for  the ``uniform'' system, Curve~2 that of a damaged ``satellite'' when all the centers are not affected. It should be pointed out that the duration of recovery process for the affected city in the ``uniform'' system turned out to be shorter than that of the ``satellite'', although the number of quanta requested by the ``satellite'' was 40~\% less than that for the city in the ``uniform'' system.  Curve~3 demonstrates that the recovery dynamics of the damaged center becomes twice as long as this process in the previous case. It is because the number of necessary quanta is much more and only the other three centers can be the donors of resources. Judging from the obtained results, the recovery process is more efficient for the ``uniform'' system than for the ``centralized'' one, if all the other factors being the same.

\begin{figure}
\begin{center}
\includegraphics[width=0.55\columnwidth]{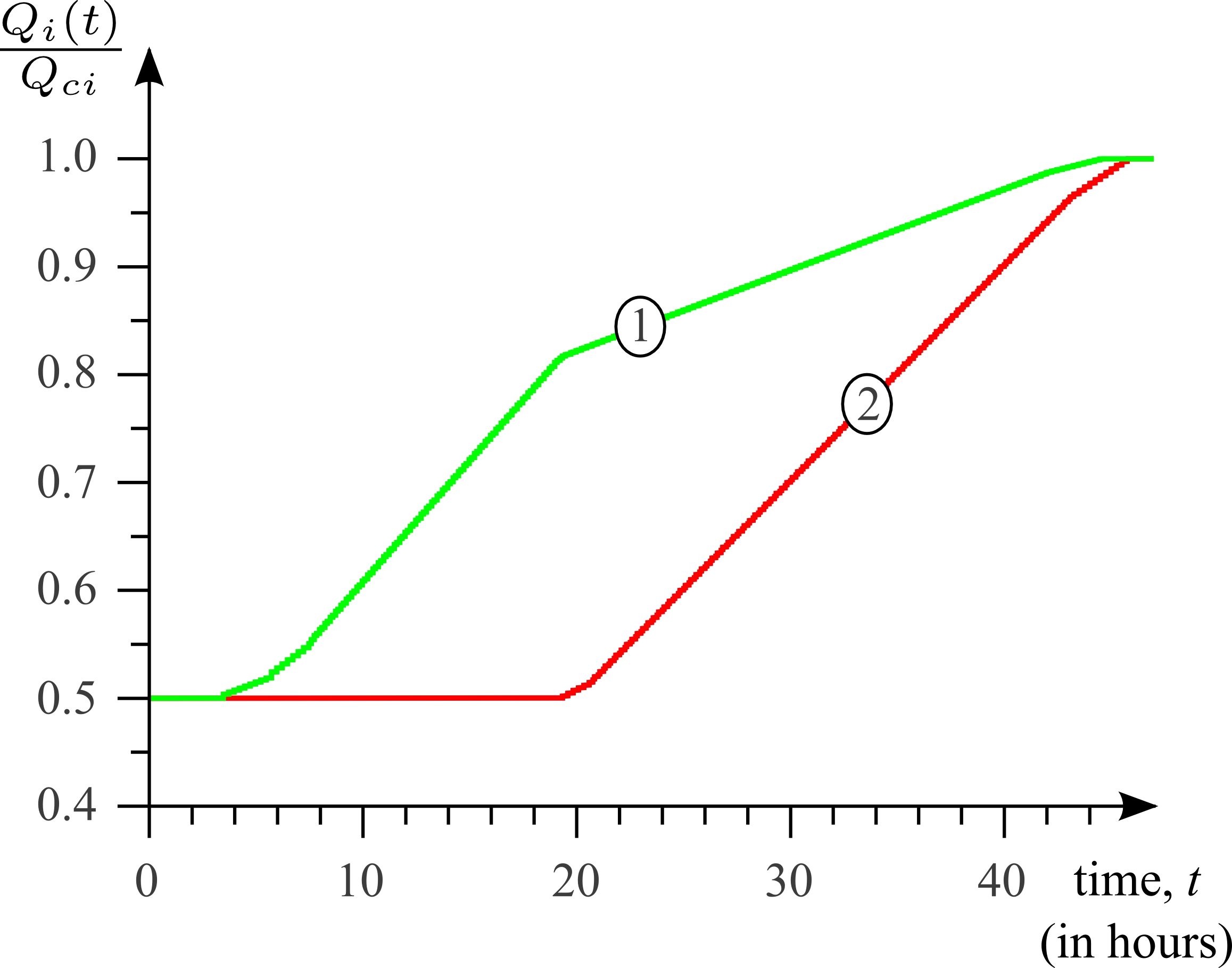}
\end{center}
\caption{The recovery dynamics in the case where the group of affected cities includes a center and ``satellites''. Curve~1 shows the recovery process of the center and Curve~2 the damaged ``satellite''.}
\label{F:Q(t)screen}
\end{figure}

The last Figure~\ref{F:Q(t)screen} illustrates an characteristic feature of the resource redistribution in the ``centralized'' system when the both types of cities (centers and satellites) are damaged. The population in the damaged center as well as $Q_c$ are much larger than those in the ``satellite'', respectively. Therefore, the priority measure $S$ of the center is also higher. It explains that the resource flow from the donors is directed to the damaged center only for a relatively long time interval. Only when the priority measures of the center and ``satellite'' becomes equal, the resource flow is shared between them. We call it the ``screening'' effect.  

Figure~\ref{F:Q(t)screen} also demonstrates a general characteristics of the resource redistribution governed by the developed algorithm. Even if the damaged cities are different in such parameters as $N_i$, $Q_i$, $Q_{ci}$, etc., the recovery process is completed practically at the same time.  It is one of the necessary properties for the algorithm to be strictly optimal. Therefore, we regard the proposed mechanism as semi-optimal. The question who it is close to the strictly optimal algorithm is worthy of individual analysis.

\subsection{Details of modeling: casewith gradually updated information}

The model under consideration is a ``uniform'' system of 81 cities with the total population $P = 8\times 10^6$; the damaged degree was set equal to $a=3$. To make the resulting dynamics of the resource redistribution smoother the limit capacity of one city $c_i$ and the time $\Delta t$ required for formating one resource quantum were chosen equal to $c_i= 1$ and $\Delta t =1$~min, which is approximately equivalent to the case of $c_i=15$ and $\Delta t = 15$~min studied previously. The other details are the same and given in Sec.~\ref{sec:det_availinf}.    

\subsection{Resource redistribution dynamics and spatial patterns}

\begin{figure}
\begin{center}
\includegraphics[width=1.0\columnwidth]{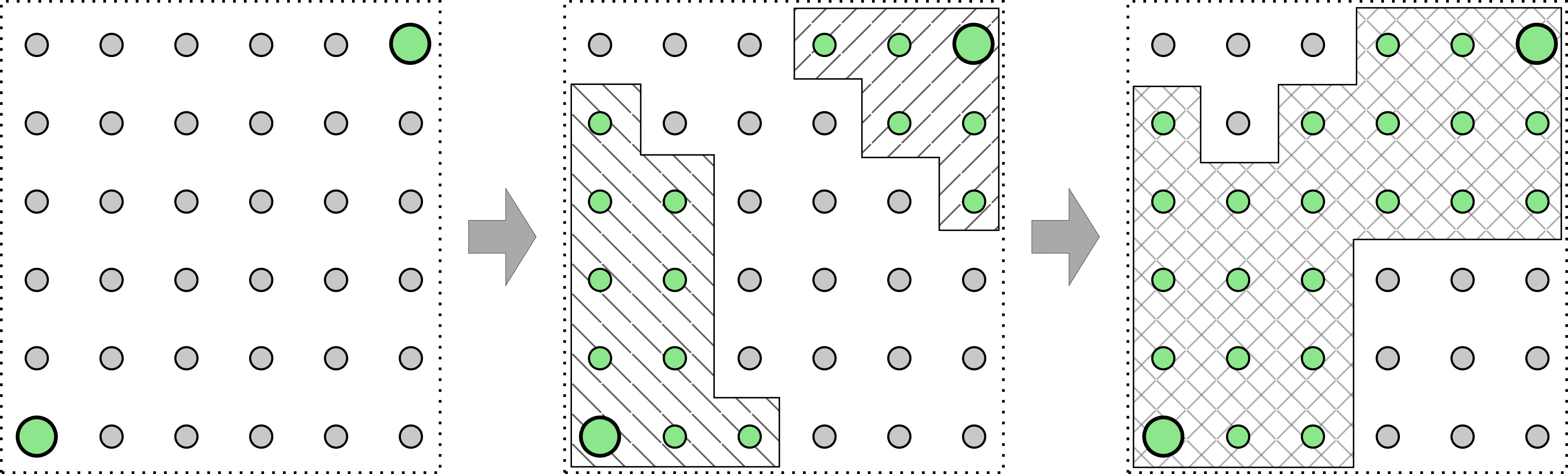}
\end{center}
\caption{Schematic illustration of a new communication system.}
\label{F:2centers}
\end{figure}

The situation under consideration is as follows. There are two individual consequences of disaster. First, nine cities were affected which cause the resource deficit there. Second, the communication network crashed. The latter means that the cities can collect only the information about their own state but their communication is not possible. Two cities located at the opposite corners are assumed to possess mobile communication facilities which can be used to create a new communication network. To do this the two cities transport the  required facilities to other cities and form two independent communication networks. Just after the relevant equipment has been delivered to a citie $i$ it is joined into the growing network and the information about its state becomes accessible for all the cities belonging to this network. Now if this newly joined city is damaged it can request for the resources or, otherwise, be included into the process of the resource supply to the affected cities previously incorporated into the same network. As soon as at least one city has joined the both networks it is assumed they are capable to exchange the collected data and,  the both networks are merged to form one network. Figure~\ref{F:2centers} illustrates this process.

\begin{figure}
\begin{center}
\includegraphics[width=0.55\columnwidth]{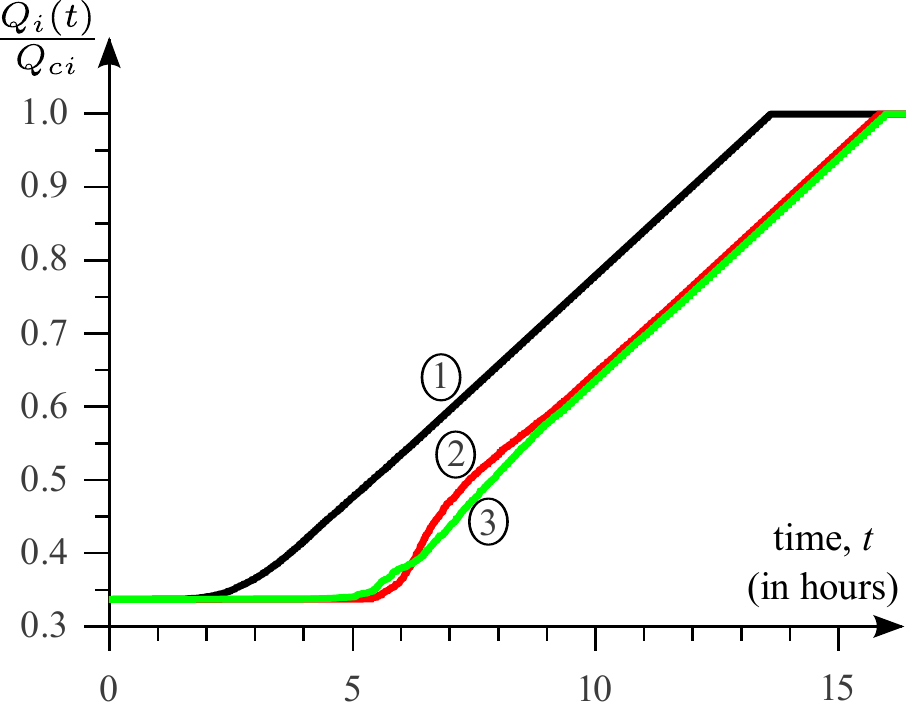}
\end{center}
\caption{The recovery dynamics when the initial communication network was ruined and later were gradually reconstructed during the recovery process simultaneously with resource redistribution. Curves~2 and 3 show this dynamics for the affected cities those were the last and the first, respectively, to be joined into the newly forming communication network. Curve~1 presents the recovery dynamics in the same system where, however, the communication network was not initially destroyed; it is presented to simplify understanding the effects of uncertainty.}  
\label{F:new1}
\end{figure}

As noted above, 9 of 81 cities located at the system center were affected by disaster and their critical level of the vital resources becomes three time as large as the amount of these resources available initially, $Q_{ci}^\text{aff} = 3 Q_{i}(0)$. Figure~\ref{F:new1} depicts the recover dynamics for three cities in two cases. First, to simplify understanding the effects of uncertainty Curve~1 exhibits the recovery dynamics of a damaged city $i_{C1}$ when the initial communication system has not been damaged. Curves~2 and 3 show the recovery dynamics of the affected cities, $i_{C2}$ and $i_{C3}$, those were the last and the first, respectively, to be joined into the newly formed communication network. Special attention should be paid to the following fact. The time difference between the moments when first resource quanta were delivered to cities $i_{C3}$ and $i_{C1}$ is about 2.5~hours. This time difference between the cities $i_{C2}$ and $i_{C1}$ is 3.8~hours. So it might be expected that the duration of the resource redistribution process in the case of the damaged communication network should be also 3.8~hours longer than this process represented by Curve~1. However, due to the cooperative interaction of the cities the resulting duration is about 2.5~hours longer. So we may state that the total duration of the process is determined by the time moment of finding one of the affected cities for the first time if, at least, their arrangement is not too heterogeneous in space. Moreover, the obtained result justifies the efficiency of this emergent recovery process even in the case when several centers start their individual operations independently of one another and merge their activities into one common process only at the final stage. 

\begin{figure}
\begin{center}
\includegraphics[width=1.0\columnwidth]{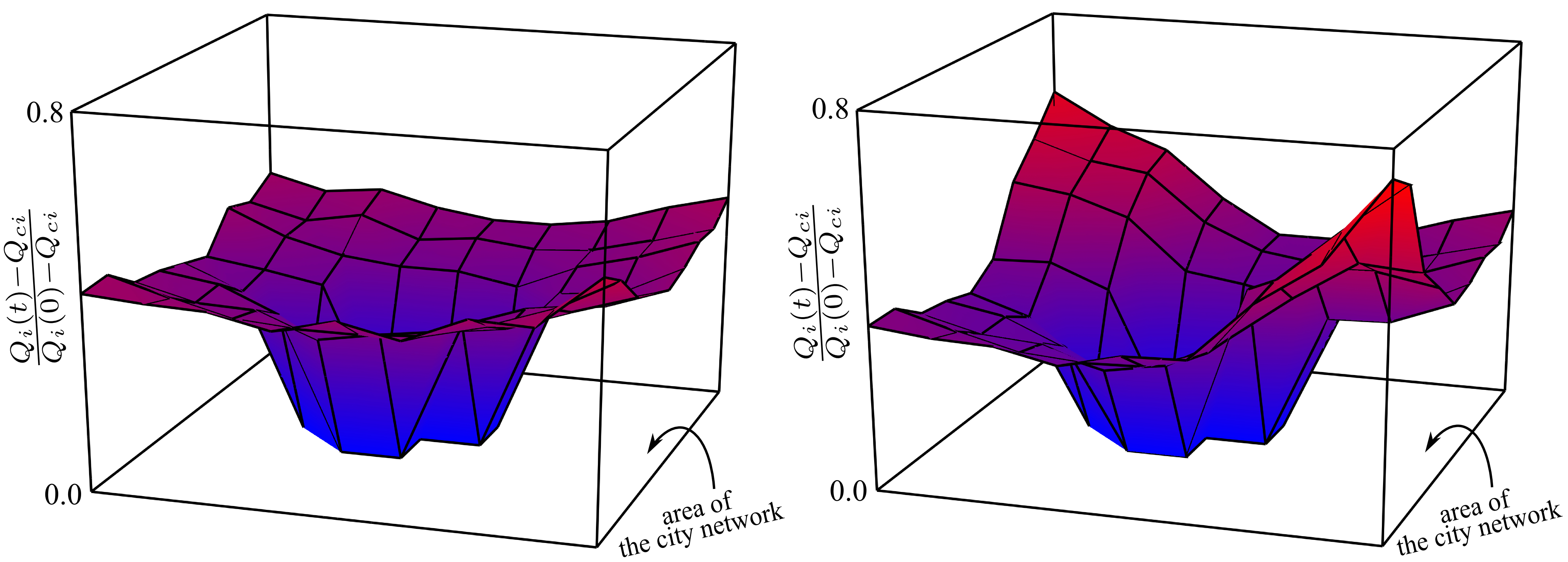}
\end{center}
\caption{The spatial pattern of the extra volume of resources distributed in the system after the recovery process has finished for two different initial conditions of a communication network. The left frame corresponds to the recovery process with  initially available information about whole system state, the right plot represent the spatial pattern for a process which was limited by the growth of newly forming communication network. As clearly shown, the right frame is highly nonuniform in the unaffected region.}
\label{F:new2}
\end{figure}

Figure~\ref{F:new2} shows the spatial patterns of the extra volume of resources distributed in the system after the recovery process has finished. The left frame visualizes this pattern in the case when the initial communication networks was not damaged. The right frame corresponds to the opposite case where involving the undamaged cities into the resource redistribution was limited by the growth of newly formed communication network. As seen, it makes this spatial pattern highly nonuniform in the unaffected region. 

The analysis of the explored model (Fig.~\ref{F:2centers}) demonstrates that the approach developed in the present paper is more appropriate than the conventional methods based on linear optimization. The matter is that during a rather long time interval the information about the system state is incomplete. Therefore, on one hand, a plan generated by the linear optimization technique cannot be strictly optimal because it depends on wrong data about the system. On the other hand, the convential approach ignores the ethic aspects represented by the triage principle. The developed approach takes into account directly the ethic factors and provides semi-optimal plan.

\section{Conclusion}

The short-term recovery of a region damaged by a large scale disaster has been under consideration. The short-term recovery  can be represented as a collection of actions with the common goal of restoring the corresponding life-support system to the minimal operating standards. The implementation of these actions could require a sufficiently large amount of resources (pure water, food, medical drugs, fuel, etc.) that are not available in the affected cities. Therefore, the resource redistribution of the vital resources becomes one of the key tasks of the short-term recovery.    

The present paper has proposed a certain method by which a plan of the required resource redistribution can be constructed. This method is based on two criteria. One of them is the optimization of the recovery duration. The other being of ethical nature is based on the triage principle. For these reasons this plan is called semi-optimal. Its features are as follows. First, since this plan is created via a certain algorithm using the data collected after the outbreak of disaster, it is not based on any pre-planning. It is suitable, because the location, time, and consequences of the disaster are unpredictable within the required accuracy and detail. Second, the corresponding resource redistribution is a decentralized process in that there are no predetermined centers through which the main part of resource flow passes and is governed by it. Naturally the headquarter is responsible for the collection of information, its processing, and acceptance of the generated plan for implementation. Thereby we imply that the process implementation is decentralized whereas its management could be centralized. The decentralized resource redistribution enables the system to react to a disaster practically immediately and makes the recovery process cooperative. Due to the cooperative effects the size of the region involved in the recovery process becomes controllable.      

The proposed algorithm includes the following. Each city $i$ is characterized by the initial amount $Q_i$ of vital resources, its critical level $Q_{ci}$ and the population $N_i$. As a result of disaster, the critical level in the affected  cities is assumed to exceed the initial amount of resources, $Q_{ci}>Q_i$; in the other cities the opposite condition $Q_{ci}<Q_i$ holds.  The key point of the developed algorithm is how to deliver the required amount of vital resources to the affected cities from the neighboring ones in a certain semi-optimal way minimizing the duration of the recovery process. To measure the lack of resources in a given city, the quantity $\theta_i=\dfrac{Q_{i}-Q_{ci}}{Q_{ci}}$ has been introduces and the value $S_i=\theta_iN_i$ has been used to order the damaged city according to the priority of resources to be received. The cities are also characterized by the limit capacity of preparing and sending quanta of resources. Exactly this limit capacity endows the resource redistribution process with nonlinear properties. The matter is that when the limit capacity is attained, the ability of a city to send a new quantum is depressed for the time necessary to prepare it. The developed algorithm simulates this effect via temporal renormalization of the real time distances between the cities.  

The main attention has been focused on the recovery dynamics for the ``uniform'' system studied numerically. In particular, it has been demonstrated that a significant growth of the degree of damage matches much weaker increase in the duration of recovery process. It is due to the cooperative effects in the resource redistribution, namely, the higher the damage level, the more the number of cities involved in the resource delivery. Second, the city limit capacity is a controllable characteristic of the system that can affect the size of the region involved in the resource redistribution as well as the portion of these cities whose state drops to the minimal operation conditions after the recovery process has finished. 

The constructed model uses the notion of administrative unit that can implement the short-recovery process on its own. The conducted numerical simulation demonstrated that for each particular situation the dependence of recovery duration on the number of cities that can be involved in the resource redistribution exhibits saturation as this number increases. Actually it specifies the dimensions of the most appropriate administrative units that are to be involved in the disaster mitigation.

The recovery dynamics in the ``uniform'' and ``centralized'' systems has been compared. The latter system was assumed to contain just four big centers able to supply the surrounding ``satellites'' with vital resources. It has been demonstrated that in this case the cooperative effects are depressed giving rise to an increase in the recovery duration. If one of these centers is affected, the duration of the recovery process increases drastically. 

Besides, as found out in the case where the center and ``satellites'' are affected simultaneously, the individual recovery processes finish for all the cities practically at the same time in spite of the difference of the cities in size, population, and the required amount of vital resources. It is one of the necessary feature for an algorithm to be optimal and allows us to call the proposed recovery plan semi-optimal.         

The effect of information uncertainty on the resource redistribution during the short-time recovery has been studied for the case when the initial communication network crashed due to the disaster and a new one is constructed gradually during the recovery process. By way of example, it has been assumed that there are two distant centers having communication facilities. A new communication network is gradually created via transporting these facilities to the cities, giving rise to updating continuously the information about the system state; the resource redistribution is implemented simultaneously with the information update. It shows that due to the cooperative interaction of the cities in redistributing the resources, the duration of such a recovery process becomes longer for the time interval required for finding one of the affected cities for the first time, at least, if their spatial arrangement is not too heterogeneous. We also have drawn a conclusion that the efficiency of this emergent recovery process is not decreased remarkably if at the initial stage several subsystems operate independently of one another and the activities merge only at the final stage.

\subsection*{Comments concerning multiple types of resources}

The present paper has considered the redistribution of single type resources only. Nevertheless the developed approach can be directly generalized to the case of multiple types of resources. If resources of different types are allocated in individual warehouses, then their redistribution processes can be implemented independently of one another and the proposed approach should be just applied in parallel to each of them. When these resources are stored in the same warehouses, due to the warehouse limit capacity their redistribution processes can interferer. In order to tackle this problem we have to introduce new variables $\{\theta_{i\alpha}\}$ in a way similar to expression~\eqref{eq:theta}, each of them quantifies the lack of the corresponding type $\alpha$ resources in a given city $i$ in relative units. At the next step taking into account the detailed mechanisms of recovery process we should construct, first, a function $S_i = S_i\{\theta_{i\alpha}\}$ evaluating the cumulative priority of the given city as a whole in the resource supply with respect to the other cities.  Second, it is necessary to construct a collections of functions $\sigma_{i\alpha}= \sigma_{i\alpha}\{\theta_{i\alpha'}\} $ specifying the priority of the individual resources in a given city $i$. We note that in the general case each function $\sigma_{i\alpha}\{\theta_{i\alpha'}\}$ includes all the quantities $\{\theta_{i\alpha'}\}$ in the list of its variables.  In these terms the cumulative priority functions $S_i\{\theta_{i\alpha}\}$ specifies the city to which the resources will be sent at each step whereas the relative priority functions $\sigma_{i\alpha}\{\theta_{i\alpha'}\}$ determine which type of resources will be sent.












\end{document}